# Tunable kHz distributed feedback fiber laser enabled by glass additive-manufacturing

**PAWEL MANIEWSKI,**[1,2,*] **ALEX I. FLINT,**[2]
**REX H. S. BANNERMAN,**[2] **TIMOTHY LEE,**[2] **MARTYNAS BERESNA**[2]

[1]*Department of Applied Physics, KTH Royal Institute of Technology, Stockholm, Sweden*
[2]*Optoelectronic Research Centre, University of Southampton, United Kingdom*
*\*pawelma@kth.se*

**For short sections of fiber tailored to a specific application, fast laser-based manufacturing techniques can be considered as an attractive alternative to the often-cumbersome traditional manufacturing routes. With the use of high-power lasers, localized hot zones that are necessary for glass making can be obtained rapidly. For instance, Laser-Powder-Deposition enables rapid fabrication of short, high gain fibers used in e.g., distributed feedback fiber lasers (DFFL). DFFLs offer sought after performance suitable for a broad range of applications in modern photonics i.e., superior stability and narrower, single-frequency linewidth compared to conventional fiber lasers. Tunable, narrow laser sources with output in eye-safe spectrum are desired for sensing, signal multiplexing, LIDAR systems, quantum applications etc. In this work we present DFFL obtained using Laser-Powder-Deposition made Er-doped silica fiber. Milliwatt level, narrow line lasing (< 704 kHz, equipment limited) was obtained using a phase-shifted grating written in 16 mm long fiber. The backward slope efficiency was as high as 24% when pumping at 976 nm. The results presented in this work showcase new possibilities in fiber fabrication that were unlocked through laser-assisted additive manufacturing. This fiber laser sets the stage for the future of rapid fabrication of advanced fiber devices through unconventional manufacturing routes.**

## 1. Introduction

Photonics devices based on silica fibers are highly regarded for their low transmission loss, robustness and easy integration with the existing infrastructure. The physicochemical properties of silica glass make these fibers suitable to work in e.g., harsh environments [1–3]. These properties can be modified by doping the glass network with other elements. For instance, by doping silica with $Ge^{4+}$ or $Al^{3+}$ its refractive index can be increased. A doped core combined with pure silica cladding is often used to obtain the typical step-index profile of the optical fiber. Furthermore, doping the silica fiber core with rare-earth (RE), enables optical gain in relatively broad spectral range [4]. For narrow linewidth emission selective feedback can be used, such as the reflection from a fiber Bragg grating (FBG). To obtain the grating, UV lasers or pulsed high peak power i.e., femtosecond lasers (femtosecond laser writing - FLW) are typically used [5–7]. In the case of UV-based writing, photosensitivity of the fiber is required. The latter can be obtained through doping with germanium (during fiber fabrication) [8] or hydrogen loading (post-processing) [9]. In the case of FLW, photosensitivity is not required, as modification is created using nonlinear absorption responsible for triggering initial seed electrons [10]. Furthermore, type II gratings, i.e., creating periodic damage in the fiber, can be preferred for some applications due to their long-term stability and robustness.

The distributed feedback fiber laser (DFFL) is a special type of fiber laser that utilizes a periodic modulation of refractive index within the active fiber to provide both gain and feedback for lasing.

Typically, to obtain the feedback a phase-shifted Bragg grating is written directly into the core of an active fiber, and thus acts as a distributed reflector [11–13]. The phase-shift introduces a localized resonance and only allows operation in a single longitudinal mode [14]. Such grating structures typically produces a stable, narrow linewidth output desirable in many applications. For instance, in telecoms, narrow linewidth sources within C-band (1530 – 1565 nm) enhance the performance, reliability, and capacity of telecommunication systems by enabling precise modulation and reducing signal cross-talk [15–17]. Gain in C-band is often obtained through Er-doping and typical pumping schemes at 980 nm or 1480 nm [14].

Generally, to obtain a high gain per unit length, a high doping concentration is desired. However, in the standard Er-doped fiber fabrication processes e.g., using modified chemical vapor deposition, dopant concentration is often limited by the Er-ion clustering leading to an increasingly efficient cooperative up-conversion processes. Both cause the so-called concentration quenching of the optical gain [18,19]. Alternatively, laser powder deposition (LPD) can be used to obtain silica-based glass rods with desired doping. The high deposition rate of > 1 $mm^3$/s enables the complete fabrication of glass rods in under a minute. The printed glass rods are then used as components for fiber manufacturing, similar to rod-in-tube technique. In LPD, a mid-IR $CO_2$ laser is used to locally melt, and sinter oxide powders delivered into the beam via off-axis jets [20]. Leveraging this additive manufacturing (AM) approach enables working with smaller production batch sizes, decreasing length of each production cycle to expedite material optimization [4]. Furthermore, in RE-doped specialty fiber manufacturing LPD enables high doping concentrations with low clustering [21]. Low clustering combined with high doping allows utilizing shorter fibers without sacrificing the gain. Consequently, short fibers mitigate undesirable nonlinear effects in the cavity e.g., stimulated Brillouin scattering or stimulated Raman scattering [22,23].

In this work, we showcase a tunable DFFL that was enabled by the combination of short, LPD-made Er-doped fiber and direct laser writing. To obtain the fiber, an Er/Al co-doped silica rod was printed using an oxide nano-powder mixture. The rod was then sleeved in a quartz tube and then the assembly was drawn into a fiber. Due to the low UV-sensitivity of the fiber, the DFFL structure was written in 16 mm long fiber using FLW plane-by-plane writing approach [11]. The lasing was obtained with a low threshold of 1.29 mW and high backwards slope efficiency up to 24 %. A narrow linewidth, less than 704 kHz (equipment limited) was obtained through heterodyne measurement. Furthermore, high stability of up to 2% power variation was measured at maximum output power. The laser output was temperature sensitive, with approximately 11 pm/K spectral shift obtained when heating up the DFFL up to 345 K. The results presented in this work showcase new possibilities in fiber fabrication that were unlocked through laser-assisted AM. This narrow line fiber laser with precise tunability with temperature sets the stage for the future of rapid fabrication of advanced fiber devices.

2. Results

To obtain the laser, first a cylindrical rod was produced through laser powder deposition. A mid-IR $CO_2$ laser ($\lambda$ = 10.6 μm) was used to sinter a nanopowder mixture as described in the procedure of rapid fabrication of silica specialty fibers [24]. The rod was subsequently sleeved in a quartz tube to form a preform with outside diameter of 6 mm, and then it was drawn into a fiber that had approximately 125 μm outer diameter. The fiber had an effective refractive index of the core mode of approximately n ≈ 1.4484 (at $\lambda$ = 1550 nm), while the transmission loss of the fiber at $\lambda_p$ = 976 nm was approximately 14.8 dB/m (see Methods). The drawn fiber was cleaved using a standard cleaver equipped with diamond blade, and then spliced to commercial single mode fiber.

In this work, a third order Bragg grating was written to ensure high contrast of the grating pitch and uniformity of the grating. To obtain the grating, plane-by-plane FLW approach was used [25] i.e., the DFFL structure was inscribed by translation of the fiber in respect to the stationary beam across and along its axis as sketched in Figure 1. The DFFL structure can be simply thought of as two identical gratings separated by a single π-phase shift. During writing the core was irradiated only during the scans across the core (100 μm/s), while each consecutive plane was offset using translation at low speed (50 μm/s) to avoid errors produced by acceleration and deceleration of the stage. In other words, the slow scans allowed precisely control of the position of each plane of the grating. In the FLW system a dichroic mirror was placed above the focusing objective and it was used to monitor the writing process in real-time using a CMOS camera. This monitoring system also allows for positioning of the fiber with respect to the laser focal point with sub-micron precision.

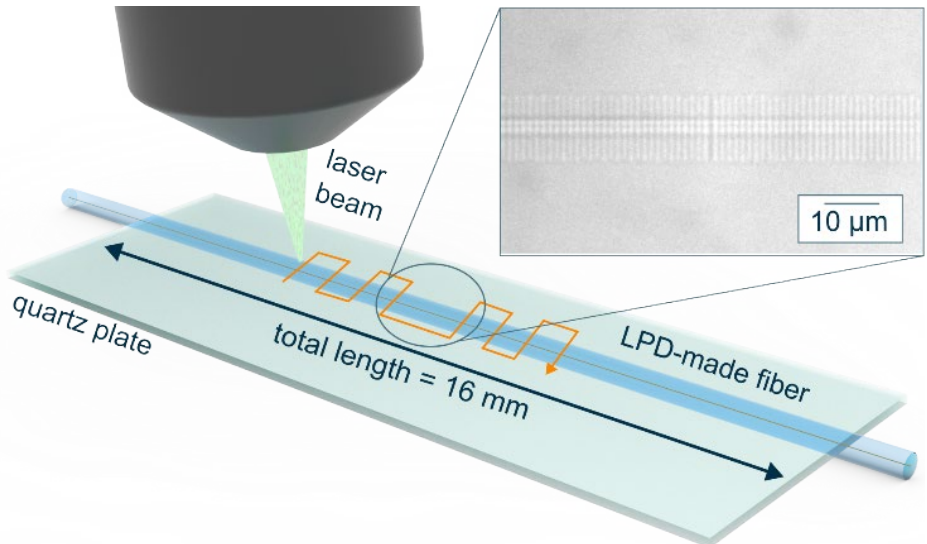

Figure 1 Schematic of fs-laser writing of DFFL. Inset shows micrograph of fiber core with fabricated grating (contrast increased for clarity).

For the writing, the 16 mm long LPD-obtained fiber, was mounted onto a quartz plate under slight tension. To compensate for slight misalignment of the fiber the start and end points of the grating were mapped in three-dimensions prior writing. This step also ensured optimal distribution of the grating throughout the whole length of the active fiber. The total length of the used gain fiber (and the grating) was 16 mm with a centrally located phase-shift. To monitor grating inscription, a pump diode laser ($\lambda_p$) was coupled into the fiber via a wavelength-division-multiplexer (WDM) and an optical spectrum analyzer (OSA) was used to monitor the signal in reflection. Such an arrangement enabled observation of how the inscribed grating was affecting the amplified spontaneous emission signal and to confirm lasing after DFFL structure was completed.

To assess the performance of the fiber and the cavity, a simple test setup was used in order to reduce the impact of unrelated factors. A schematic of DFFL experimental setup is shown in Figure 2. The forward output face was terminated in order to restrain any back reflections. Using a laser diode ($\lambda_p$ = 976 nm) as a pump, a lasing threshold of 1.29 mW was obtained. Backwards slope efficiency of up to 24 % was measured against absorbed pump (see Figure 3). It is noteworthy that due to the simple symmetric-grating design utilized in this study, DFFL was lasing in both directions. Our focus on this straightforward design allows for a clear demonstration of the lasing performance, while more complex cavity designs are well-documented in the literature and can be utilized in our future work. Power stability was measured and a variation of up to 2% was found when running at maximum power for over 1 hour.

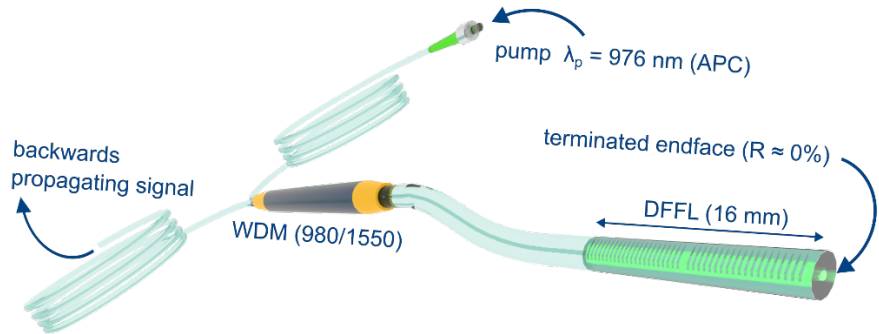

Figure 2 Schematic of DFFL experimental setup.

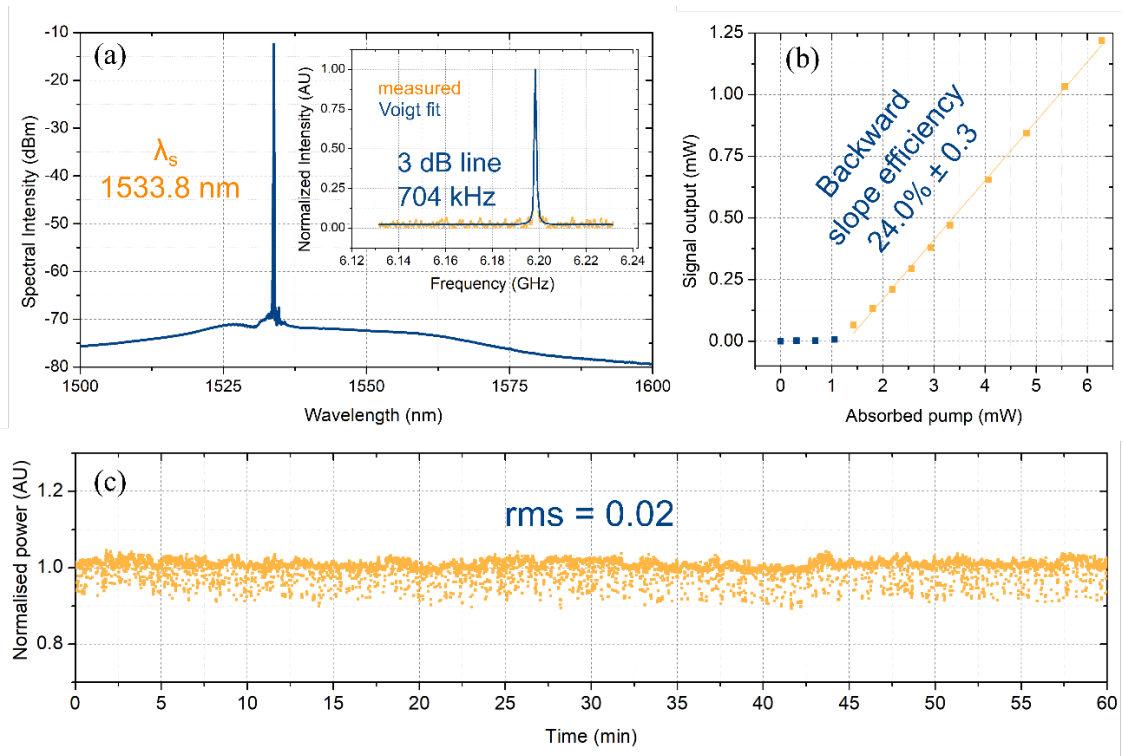

Figure 3 DFFL output spectrum obtained readily post-FLW (a) with a normalized heterodyne beat note with a local oscillator (inset) and its backward slope efficiency (b) against the absorbed pump. Normalized to average power stability (c) at maximum output.

An upper limit of the linewidth was determined by heterodyne measurement. The 3 dB linewidth was found to be 704 kHz as shown in Figure 3 (a). The beat note in such a measurement is larger than the linewidth of the lasers used [26,27], and therefore, the true linewidth is expected to be narrower.

Thermal shift of the output was investigated at elevated temperatures as shown in Figure 4. To assess the shift of the laser, it was mounted onto a silicone heating mat and placed in an insulated box. The output was monitored using the OSA. The temperature offset was measured with a thermocouple placed in close proximity of the fiber. A shift of approximately 11 pm/K was observed in temperatures up to 345 K.

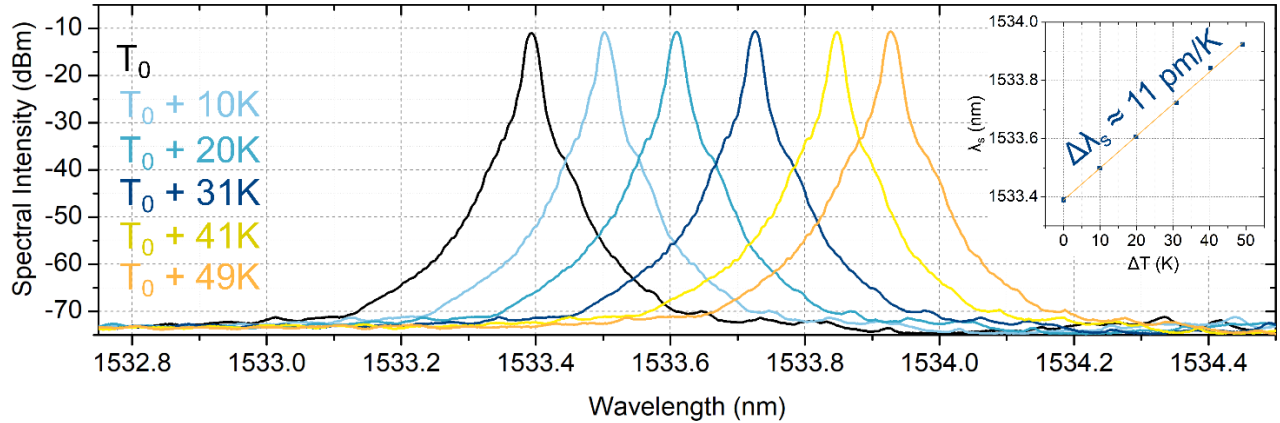

Figure 4  Spectral intensity of the DFFL's output at different temperatures ($T_0$ – room temperature). Inset shows shift of the laser peak, that corresponds to 11 pm/K.

In summary, in this work we demonstrated to the best of our knowledge, the first narrow linewidth fiber laser that was based on short, active fiber obtained through additive manufacturing. Here, selective, distributed feedback was obtained using a femtosecond laser direct writing method. This approach reduces fabrication time significantly while also achieving enhanced performance of the device. Demonstrated here is eye-safe, milliwatt output, showing high stability and high efficiency. The high efficiency of up to 24% of the laser, only measured backwards, is one of the highest reported for this type of material, doping and pump scheme [11,12,28]. Further improvements that include e.g., co-doping with $Yb^{3+}$ to improve pump absorption [14,15,29] or more complex DFFL cavity structures can be included in our future work. Nonetheless, the results presented in this paper underline the high quality of the LPD-obtained fiber and its suitability for DFFLs. The results represent a significant advancement in unconventional fiber laser manufacturing and thus rapid prototyping of future fiber sources. It also showcases new possibilities for swift production of tailored, high-performance devices that can be used in various high-tech fields.

3.  Methods

**Laser powder deposition**
A dry, nano-powder mixture of $SiO_2$, $Al_2O_3$ and $Er_2O_3$. (all: US Research Nanomaterials, Inc) was prepared using custom-made stainless-steel ball-mill. The ratio of powders was: 94:5.3:0.7 wt% respectively. A $CO_2$ laser (ULR-50, Universal Laser Systems Inc) was used along with a quartz plate as the substrate (flat plate, 1 mm thick, PlanOptik AG). The substrate was fixed into a two-axis stage (XY), whereas the print head was mounted onto a vertical translation stage (Z). During printing vertical translation of the print head with a speed of 1 mm/s was combined with powder feeding rate of approximately 0.5 g/min. Printing of freestanding glass rods was described in detail in [30]. The transmission loss of the resultant fiber was estimated through 12 consecutive cutbacks. The fiber loss was then calculated using an exponential fit.

## Direct laser writing

A frequency doubled (515 nm) femtosecond Yb:KGW laser system (Pharos SP, Light Conversion Ltd) operating at 200 kHz was used for grating inscription. First, the fiber was fixed onto a quartz plate and immersed in silica index-matching oil (Cargille, n = 1.4587 ($\lambda$ = 589.3 nm)). The laser beam ($M^2 < 1.2$) was focused using an oil immersion microscope objective with variable numerical aperture (Olympus Plan N 50x / 0.5-0.9) in axial plane of the fiber core. The objective is attached on vertical translation stage allowing to dynamic adjustment of the focusing position. The quartz plate was mounted via custom made sample holder onto Aerotech ANT130-160-XY stages. In Table 1 writing parameters used for the DFFL are shown. During writing, a fiber coupled laser diode ($\lambda_p$ = 976 nm, BL976, Thorlabs Inc) was coupled into the LDP-obtained fiber via a 980/1550 wavelength division multiplexer (WDM). The backward propagating output was monitored using an optical spectrum analyzer with a resolution of 0.02 nm (Yokogawa AQ6370).

Table 1. Parameters used for DFFL writing

| | |
|---|---|
| $P_{avg}$ (incident on the fiber) | 6.2 mW |
| Wavelength | 515 nm |
| Rep. rate | 200 kHz |
| Pulse energy | 30 nJ |
| Objective NA | 0.5 |
| $M^2$ | < 1.2 |
| Speed $x$ (along the fiber) | 50 μm/s |
| Speed $y$ (across the fiber) | 100 μm/s |

## DFFL evaluation

*Basic laser performance:* the fiber coupled laser diode ($\lambda_p$ = 976 nm, BL976, Thorlabs Inc) was coupled into the DFFL via a 980/1550 wavelength division multiplexer (WDM). The latter was used to collect backward propagating signal ($\lambda_s$) and monitored with an optical spectrum analyzer with a resolution of 0.02 nm, (Yokogawa AQ6370).

*Power stability:* using the above-listed layout, powder stability was measured using a photodiode (S132C, Thorlabs Inc) and collecting data for 1 hour at maximum output. A sampling rate of 2 Hz was used.

*Heterodyne measurement:* The backward propagating DFFL output was combined with a local oscillator (Agilent 81600B), using a 50:50 beamsplitter. The beat note between the two lasers was then measured. A fast photodiode (PDA8GS Thorlabs Inc) and an electrical spectrum analyzer (AVANTEST R3273) with a scanning time of 30 ms were used.

**Funding.** Swedish Research Council (2022-06180); Stiftelsen Tornspiran (894); Engineering and Physical Sciences Research council (EP/R513325/1).

**Authors contribution.** PM designed and fabricated the fiber, and conceptualized the study, AIF, RHSB and PM characterized the laser, RHSB investigated UV sensitivity of the fiber, MB, TL and PM investigated laser writing and inscribed the grating. All authors analyzed and reviewed the results and participated in writing the manuscript.

**Acknowledgements:** The authors would like to thank Peter Horak for the insightful conversations and feedback on this work.

**Data availability.** Data underlying the results presented in this paper is available in Supplement.